\documentclass[12pt,prd,nofootinbib,tightenlines,showpacs]{revtex4-1}
\usepackage{amsmath}
\usepackage{amssymb}
\usepackage{graphicx}

\begin{document}

\title{Systematics of black hole binary inspiral kicks and the slowness approximation}

\author{Richard H.~Price} \affiliation{Department of Physics \&
  Astronomy and CGWA, University of Texas at Brownsville, Brownsville
  TX 78520}

\author{Gaurav Khanna} \affiliation{Department of Physics, University
  of Massachusetts, Dartmouth, MA 02747}

\author{Scott A.~Hughes} \affiliation{Department of Physics and MIT
  Kavli Institue, MIT, 77 Massachusetts Ave., Cambridge, MA 02139}

\begin{abstract}
During the inspiral and merger of black holes, the interaction of
gravitational wave multipoles carries linear momentum away, thereby
providing an astrophysically important recoil, or ``kick'' to the
system and to the final black hole remnant. It has been found that
linear momentum during the last stage (quasinormal ringing) of the
collapse tends to provide an ``antikick'' that in some cases cancels
almost all the kick from the earlier (quasicircular inspiral)
emission. We show here that this cancellation is not due to
peculiarities of gravitational waves, black holes, or interacting
multipoles, but simply to the fact that the rotating flux of
momentum changes its intensity slowly.
We show furthermore that an understanding of the systematics of the
emission allows good estimates of the net kick for numerical
simulations started at fairly late times, and is useful for
understanding qualitatively what kinds of systems provide large and
small net kicks.
\end{abstract}

\maketitle

\section{Introduction}\label{sec:intro}

Since the breakthrough work by Pretorius\cite{breakthrough1}, and by
the Brownsville and Goddard groups\cite{breakthrough2,breakthrough3},
numerical relativity (NR) has developed to the point that at present
there are few limitations, except computer time, on the modeling of
the inspiral of binary black holes. Among other phenomena that can now
be studied numerically is the linear momentum contained in the
gravitational waves emitted during inspiral.  
The result of this emission of momentum is the recoil of the final
merged black hole, with significant consequences for scenarios of
galactic evolution and electromagnetic signals from the effect of the
recoil on the merged remnant's environment.  See, for example,
Refs.\ \cite{vgm10,zrdzp10,gmmc11,ssh11,bclh11} for recent work
discussing astrophysical implications of black hole kicks.

The process of inspiral/merger can be divided into an early slow
quasicircular inspiral, driven by gravitational wave radiation
reaction, and late quasinormal ringing at complex frequencies
characteristic of the spacetime of the final hole. The transition
between the two different regimes is often called the plunge.  

Schnittman {et al.}\ \cite{SchnittmanEtAl} seem to have been the first
to notice that NR results show a general tendency for the radiation of
linear momentum to reverse direction (although this result had been
predicted about a year earlier by Damour and
Gopakumar\cite{DamourGopakumar06}).  In deconstructing the
computations of collisions of comparable mass holes Schnittman et
al.\ found that some of the pre-plunge ``kick'' provided by the early
quasicircular inspiral is canceled by a large ``antikick'' during the
post-plunge quasinormal ringing.  This ``antikick'' can be
drastic. For one of the models studied by Schnittman et al.\, the
final total kick was only around one third of the maximum kick, a
maximum that occurs around the start of the binary plunge.

Particle perturbation methods\cite{SundararajanEtAlI,MinoBrink}, with their relative simplicity, provide
an important tool for probing points of principle, such as the
antikick cancellation, more efficiently than  full NR. Such
modeling\cite{SundararajanEtAlI} has shown that for appropriate binary
parameters (equatorial orbits, rapidly spinning background hole) the
cancellation can be almost total. A cancellation of 97\% of the
maximum linear momentum was found for an equatorial inspiral for
background spin parameter $a/M=0.9$, and mass ratio $10^{-4}$\cite{typo}. 
(The mass
ratio is relevant since it governs the rate at which the quasicircular
pre-plunge orbits decay.)

Attempts to understand this cancellation have led to a focus on the
multipole structure of the gravitational wave
emission\cite{SundararajanEtAlI,SchnittmanEtAl}.  A single multipole
carries no linear momentum. It is the interaction of modes that gives
rise to net linear momentum. The radiation in the $z=0$ plane, for
example, can carry linear momentum only if the radiation contains
modes with both even and odd azimuthal indices $m$. The linear
momentum, furthermore, is sensitive to the relative phase of the
multipoles. While this multipole analysis is important it has not
seemed to give a satisfactory answer to the underlying question: what
can it be in the conditions of the pre-plunge radiation or motion, that
``sets up'' a plunge and ringdown that provide an antikick that almost
cancels the initial kick?
Some interesting underlying mechanisms have
been proposed\cite{Rezzolla}.

We will show here that in fact the phenomenon of antikick cancellation has
nothing that is specific to gravitational radiation, to black holes,
or to mixtures of multipoles. Rather, it is a general consequence of
the way in which the radiation gradually builds up in time, and then
dies off in a way that is in some sense gradual.  Here ``gradual''
means that the  oscillatory period of the radiation is short
compared to the timescale for the change in the intensity and period
of the radiation. It turns out that this condition is not strictly
obeyed during the plunge, and that the condition is in fact quite
difficult to define, and may ultimately have to be accepted as a
qualitative criterion. Nevertheless it will be clear that this is the
basic feature of the radiation that accounts for the strong antikicks.

As already mentioned, the prediction of an antikick had already been
made in 2006 by Damour and Gopakumar\cite{DamourGopakumar06}. In
attempting to apply the effective one body approximation to kick
calculations they pointed to the importance of the ratio of the
orbital period to the timescale. It is particularly interesting that
they identified (as do we) the epoch of the plunge as the crucial time
for determining the net kick, and found that its details were crucial
and not amenable to their approximations. These conclusions are all
very similar to our own, though from a rather different point of view.

The phenomenon of the antikick, furthermore, is very robust. It
applies just as well to nonlinear models as to linearized models. This
general insight, furthermore, gives an immediate understanding of the
conditions -- at least the qualitative conditions -- under which the
antikick will give a significant reduction in recoil, and when the
antikick can be ignored. In addition, this insight provides a possible
efficiency in the computation of kicks. It shows that a binary model can be
started very late, i.e., not long before the plunge, and yet
yield a good estimate of the linear momentum radiated.

In this paper we present an analysis confined to equatorial binary
orbits (i.e., for holes whose spin, if nonzero, is perpendicular to
the orbital plane).  These are the cases in which the antikick can be
dramatically strong. We will comment only briefly and speculatively on
the extension of our analysis to more general configurations.

The remainder of this paper is organized as follows. In
Sec.~\ref{sec:equatorial} we start by giving an analysis based on a
``slow approximation.'' We introduce a simple toy model with a well
defined time scale for change, and we show that this model duplicates
the behavior found in gravitational radiation.  We then show that for
gravitational radiation computational results there is a 
correlation between the ``slowness'' of the inspiral/merger and the
extent of the antikick cancellation of the kick, but that it is
difficult to make the correlation quantitative.
 In Sec.~\ref{sec:slowapp}, we show a very different side of the slow
 approximation; we demonstrate that based on this approximation, a good
 estimate of the radiated momentum can be found from a computation started shortly
  before the plunge.  We summarize in 
Sec.~\ref{sec:conclusion} and discuss possible extensions of this work.

\section{Kick-antikick cancellation for quasicircular equatorial 
orbits}\label{sec:equatorial}

\subsection{General analysis}
We start by considering two objects in quasicircular orbits around
each other in such a way that there is a symmetry of all physical
features with respect to the orbital plane, so that linear momentum
flux must be parallel to the orbital plane.  The binary system could
be a point particle orbiting in the equatorial plane of a Kerr hole or
two comparable mass objects (black holes or stars) with their spin
axes perpendicular to the orbital plane.

We use inertial Cartesian coordinates to describe the spacetime far
from the orbiting bodies, with the orbital-symmetry plane taken to be
the $xy$ plane, and the system supposed to have its orbital angular
velocity in the positive $z$ direction. 

In general, the orbital frequency $\Omega(t)$ will be a slowly
changing function of time, and the rate of emission of linear
momentum, which we will denote $f(t)$, will be a function of time due to the
changing frequency and radius of the motion. Here all physical
quantities, the linear momentum, the frequency, and the time $t$, are
understood to be measured in the asymptotically flat spacetime far
from the binary.

The direction of emission of linear momentum will rotate at frequency
$\Omega(t)$, so the linear momentum emitted per unit time ${\bf \dot{P}}$
will have components

\begin{equation}\label{LxLy}
  \dot{P}_x=f(t)\cos{\phi(t)}\quad\quad  \dot{P}_y=f(t)\sin{\phi(t)}\,,
\end{equation}
where $\phi$ is the time-changing angle in the orbital plane between the $x$ axis
and the direction in which linear momentum is being radiated.

The total linear momentum components carried away in the waves, from
the start of emission at $t_{\rm start}$ to some particular time $t$,
are the time integrals of $\dot{P}_x$ and $\dot{P}_y$. Since $d\phi/dt$ will be
always  positive, $\phi(t)$ is an invertible function, and we can
write the time integrals as
\begin{equation}
  {P}_x=\int_{\phi(t_{\rm start})}^{\phi(t)} \frac{f}{d\phi/dt} \cos{\phi}\,d\phi
\quad\quad
  {P}_y=\int_{\phi(t_{\rm start})}^{\phi(t)}  \frac{f}{d\phi/dt} \sin{\phi}\,d\phi\,.
\end{equation}
We define $F(\phi)=f(t[\phi])/(d\phi/dt)$ and write the integrals in the suggestive form
  \begin{eqnarray}
  P_x&=&\int_{\phi(t_{\rm start})}^{\phi(t)} 
F(\phi)\cos\phi\,d\phi
=\left.F(\phi)\sin\phi+F^\prime\cos\phi-F^{\prime\prime}\sin\phi\pm\cdots
\right|_{\phi(t_{\rm start})}^{\phi(t)}\label{eq:Lx}\\
\ P_y&=&\int_{\phi(t_{\rm start})}^{\phi(t)} 
F(\phi)\sin\phi\,d\phi
=\left.-F(\phi)\cos\phi+F^\prime\sin\phi+F^{\prime\prime}\cos\phi\pm\cdots\right|_{\phi(t_{\rm start})}^{\phi(t)}\label{eq:Ly}\,.
  \end{eqnarray}
Here $F^\prime=dF/d\phi=\Omega^{-1}dF/dt$ and is of order
$(T/\tau)F$, where $ \tau$ is the timescale for change of $F$,
and where ${T}=2\pi/\Omega$ is the period of the oscillatory 
process.

For a slowly decaying binary there is a very small fractional change
in $f$ and $\Omega$, and hence in $F$, over an orbit. We tentatively
assume that this is true for any of the orbits before the plunge, and
that it is also true in the late post-plunge epoch of the merger when
$\Omega$ is to be interpreted as the frequency associated with the
quasinormal ringing of the final black hole. In short, we tentatively
assume that the ``slowness parameter'' ${T}/\tau$ is small throughout
the integration. Thus $F\gg F^\prime\gg F^{\prime\prime}\gg\cdots$,
and for slowly changing orbits keeping only the first term on the
right, or the first few terms, is a good approximation.

We now add
the assumption that the strength of gravitational emission vanishes at
the start of the binary inspiral. This is simply the statement that
the starting separation is large enough that the early emission is
tiny compared to the emission at the time of interest.
From Eqs.~(\ref{eq:Lx}) and (\ref{eq:Ly}) we then have that the total
of the momentum radiated during the inspiral from $t_{\rm start}$ to $t$ is
\begin{equation}\label{LofT}
  P_x(t)= \frac{f(t)}{\Omega(t)}\sin{\phi(t)}\left[1+{\cal O}({T/\tau})\right]
\quad\quad\quad
  P_y(t)=- \frac{f(t)}{\Omega(t)}\cos{\phi(t)}\left[1+{\cal O}({T/\tau})\right]\,.
\end{equation}
  \begin{figure}[h]
  \begin{center}
  \includegraphics[width=.9\textwidth ]{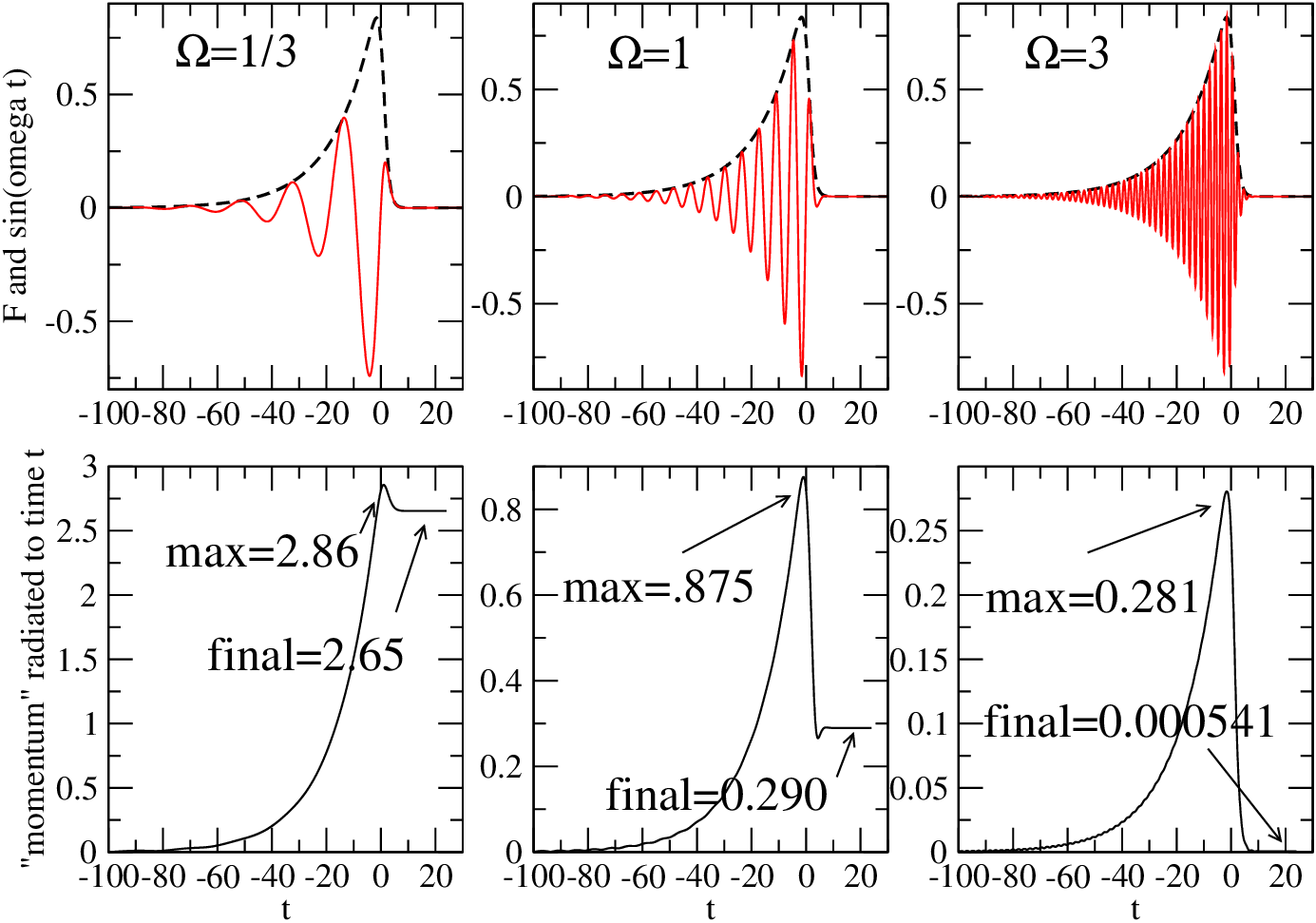}
  \caption{The top row shows the envelope function $f(t)$ for our model
problem and compares it with the oscillations $\sin(t/3)$, $\sin(t)$, and $\sin(3t)$.
The second row shows the ``momentum'' radiated up to time $t$ for models with 
the three different choices of $\Omega$.
}
  \label{fig:6graphs}
  \end{center}
  \end{figure}

For the binary inspiral of two black holes, or of a black hole and a
particle, the amplitude of gravitational wave emission increases
slowly until the start of the ``plunge'' phase of inspiral. Around
that time emission reaches a maximum, then decreases and dies out with the
damped sinusoidal pattern of the quasinormal ringing of the final hole
being formed. The epoch of plunge and quasinormal ringing is not
typically characterized by values of ${T}/\tau$ as small as those in
the slow inspiral, but in many cases, especially those involving
rapidly rotating holes, the value of ${ T}/\tau$ remains generally
small (though this issue will be more closely examined below).

We separate the binary process into the epoch $-\infty<t<t_{\rm plunge}$
before the smooth inspiral begins the inward plunge and  the
post-plunge epoch $t_{\rm plunge}<t<\infty$.  We tentatively assume that the
same ${T}/\tau\ll1$ approximation can be applied throughout the
entire inspiral, plunge and quasinormal ringing. 
Aside from corrections of
fractional order ${\cal O}({T}/\tau)$,
we have then from
Eq.~(\ref{LofT}) that the momentum emitted up to the time of plunge is
\begin{equation}\label{LofT2}
  P_x= \frac{f(t_{\rm plunge})}{\Omega(t_{\rm plunge})}\sin{\phi(t_{\rm plunge})}
\quad\quad\quad
  P_y =- \frac{f(t_{\rm plunge})}{\Omega(t_{\rm plunge})}\cos{\phi(t_{\rm plunge})}\,.
\end{equation}
Since $f$ vanishes at $+\infty$ as well as $-\infty$,  
Eqs.~(\ref{eq:Lx}),(\ref{eq:Ly}), with $t_{\rm start},t$ replaced
by $t_{\rm plunge},+\infty$ tell us that the post-plunge components are
\begin{equation}\label{LofT3}
  P_x= -\frac{f(t_{\rm plunge})}{\Omega(t_{\rm plunge})}\sin{\phi(t_{\rm plunge})}
\quad\quad\quad
  P_y= \frac{f(t_{\rm plunge})}{\Omega(t_{\rm plunge})}\cos{\phi(t_{\rm plunge})}\,.
\end{equation}

The conclusion is that the post-plunge emission cancels the pre-plunge
emission, aside from corrections of order ${T}/\tau$. It should be
noted that $t_{\rm plunge}$ can be chosen to be any value of $t$, and
so that our conclusion is that for a slow process the total momentum
radiated is negligible.  That conclusion is immediate if we put
$-\infty,+\infty$ in place of $t_{\rm start},t$ in the integrals in
Eq.~(\ref{LofT}). With this substitution it can also be seen that the
components of the total momentum emsission are equivalent to the
Fourier transform of the function $F(\phi)$ at ``frequency'' unity. In
the case that ${T}/\tau\ll1$, the function $F(\phi)$ changes little
for $\Delta\phi=2\pi$, and this Fourier component will be small.
Somewhat loosely speaking, this is the statement that antikick cancels
the kick to the extent that the rate of change of the emission is slow 
compared to the rate of orbiting, plunging and ringing.

Our analysis in Eqs.~(\ref{LofT2}) and (\ref{LofT3}) is more than just
the statement of negligible total radiation. If $t_{\rm plunge}$ is
the time at which $f(t)/\Omega(t)$ is a maximum, than those equations
imply that the momentum radiated up to time $t_{\rm plunge}$ is much
larger than the momentum that remains after the cancelation. 
If the approximation ${T}/\tau\ll1$ were strictly valid for $-\infty<t<\infty$,
then the conclusion would be that the total momentum radiated is smaller 
than the maximum by a factor of order ${T}/\tau$.

\subsection{Examples}

Before applying the above ideas to actual binary computations, it is
useful to start with the definitiveness of a simple toy model. For the
model to be simple we will choose $\Omega$ to be a constant, and will
choose $f(t)$, the analog of the ``intensity'' of $|\dot{\bf P}|$  to be the function
\begin{equation}\label{envelope}
  f(t)=\frac{e^{t/15}}{1+e^{t-1}}\,.
\end{equation}
This  function, shown as the dashed curve in the top row of the panels in
Fig.~\ref{fig:6graphs}, plays the role of  the envelope of the oscillations of 
$\dot{P}_x$ and $\dot{P}_y$. The particular choice in Eq.~(\ref{envelope})
has some flavor of the actual envelope of the
momentum oscillations for a binary inspiral. It starts slow, at large
negative times, with $f(t)\approx e^{t/15}$, and finishes at a more
rapid pace, with $f(t)\propto e^{-14t/15}$.  We can take the timescale
for change to be $\tau=|f/(df/dt)|$. This timescale starts at $15$ at
large negative times, diverges at around $t=-1.64$ and approaches a
magnitude  $15/14$ at large positive times.  For this simple envelope
the minimum timescale is therefore 15/14. 

The analog of $\dot{P}_y$ is $f(t)\sin{\Omega t}$ where $\Omega$ is the 
orbital/ringing frequency (constant for our toy model). We consider
three possible values  $\Omega=1/3$, 1, or 3. (With a simple rescaling
$t^\prime=\Omega t$
this is equivalent to the choice of a single orbital/ringing frequency
and three different ``slownesses'' of the envelope
function $f(t)$.)
With periods ${T}=6\pi, 2\pi, 2\pi/3$, the slowness parameter ${T}/\tau$ in
each case has a maximum of 17.6, 5.86, and 1.95, respectively. In the
top row of Fig.~\ref{fig:6graphs} the envelope is shown along with the
sinusoidal oscillations. It clear that these numerical indicators are
compatible with the visual appearance of the curves. For $\Omega=1/3$,
the envelope (at large time) changes more rapidly than the
oscillations; for $\Omega =1$ the envelope and oscillations change at
a comparable rate; for $\Omega =3$ the envelope changes more slowly
than the very rapid oscillations.

The ``momentum'' radiated in the toy model has been computed for each of the three 
cases. That is, the following integrals have been computed
\begin{equation}\label{toyPs}
  P_x(t)=\int_{-\infty}^{t} \frac{e^{t^\prime/15}
\sin{\Omega t^\prime}}{1+e^{t^\prime-1}}\,dt^\prime
\quad\quad
  P_y(t)=\int_{-\infty}^{t} \frac{e^{t^\prime/15}
\cos{\Omega t^\prime}}{1+e^{t^\prime-1}}\,dt^\prime\,.
\end{equation}
The results for the total momentum radiated $\sqrt{P_x^2+P_y^2\;}$\,, as a
function of time, are shown in the second row of of
Fig.~\ref{fig:6graphs}.  In the case of the truly slow envelope, with
$\Omega=3$, the ``kick'' (maximum of the
momentum radiated) is 0.281, while the final value .000541 is only
0.2\% of the maximum kick. The post-``plunge'' antikick has canceled
99.8\% of the pre-``plunge'' maximum. The remaining cases, for
$\Omega=1$ and 1/3, show that there is a clear correlation of the 
slowness and the extent to which the antikick cancels
the kick. They also show that while the correlation is strong, it is 
not simple. The ratios of final momentum to maximum momentum is 
0.93, 0.33, 0.0019 for $\Omega=$ 1/3, 1, 3 respectively. This is a much
more dramatic dependence on slowness than the ratios of the timescales.

\begin{figure}[h]
  \begin{center}
  \includegraphics[width=.7\textwidth ]{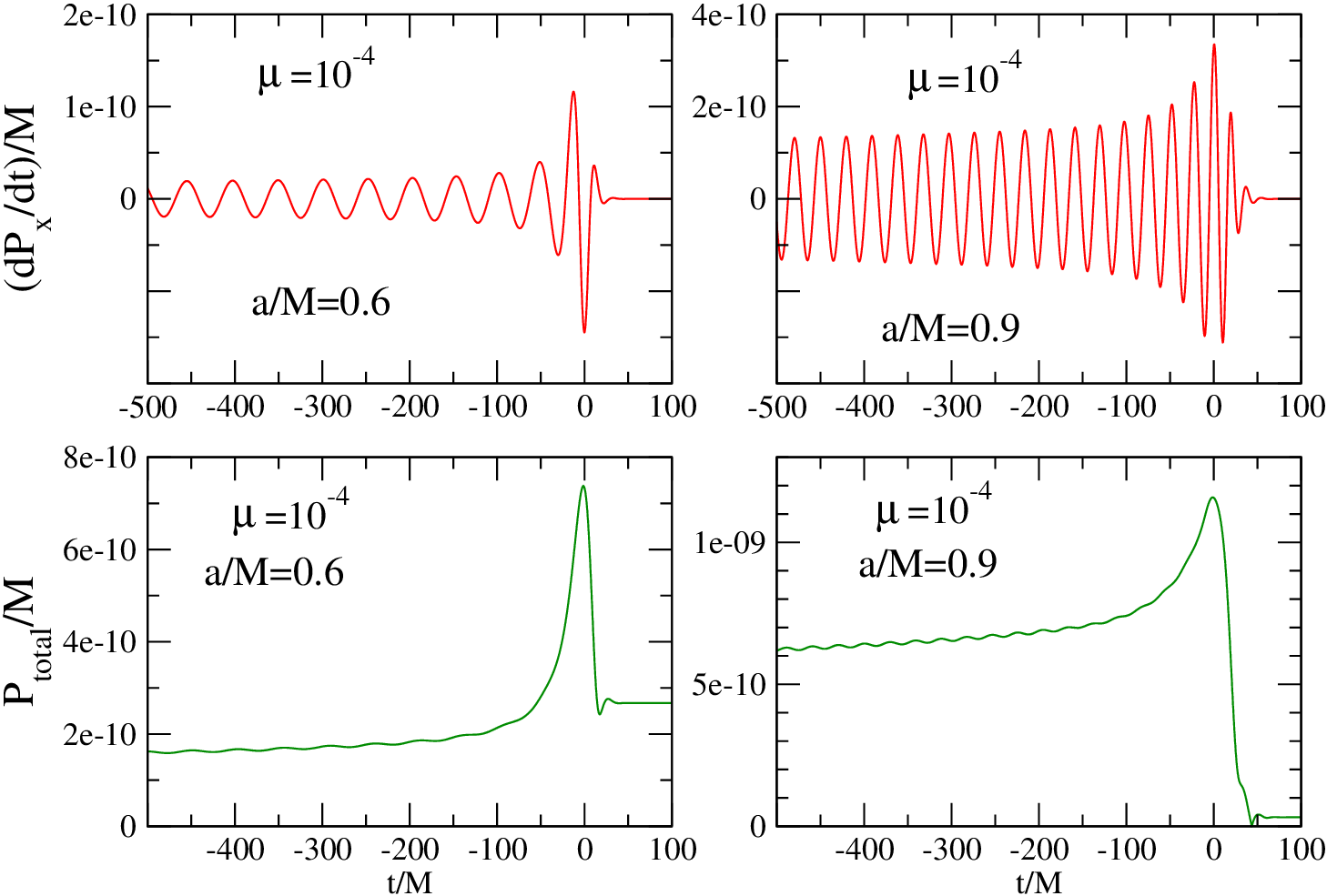}
  \caption{Computational results for mass ratio $\mu=10^{-4}$ in the case of 
black hole spin parameters $a/M=0.6$ and $a/M=0.9$.}
  \label{fig:mu-4}
  \end{center}
  \end{figure}

We now look at similar considerations for actual particle perturbation
models. We start with the results that were used in
Ref.~\cite{SundararajanEtAlI}, and that correspond to a very small
mass ratio (particle mass to black hole mass) $\mu=10^{-4}$, and hence
to a very slow early pre-plunge epoch, when inspiral is driven by the
loss of orbital energy to gravitational waves. The plots in the top
row of Fig.~\ref{fig:mu-4} show $\dot{P}_x$ as a function of time. 
The particular values of $t/M$ have no absolute meaning, so we have
translated the results here, and in subsequent plots, so that $t/M=0$ 
always corresponds to the time of maximum radiated momentum.

At
early time these oscillations are characteristic of the rotation of
the system. (In this particle perturbation model it is the rotation of
the particle in the Kerr background.) The intensity of the
gravitational raditation increases until the sytem reaches its
smallest stable separation, at around time $t/M=0$. This point of maximum emission
roughly signals the start of the plunge as the particle goes inside
the radius for the innermost stable circular orbit. The subsequent
radiation pattern, within an oscillation or two, becomes that of
quasinormal ringing.

A striking feature  of Fig.~\ref{fig:mu-4}
is that the  frequency for the pre-plunge momentum oscillations 
for $a/M=0.9$ is significantly higher than for $a/M=0.6$. This is
due to the fact that the radius of innermost
stable circular orbit, the ``ISCO'' in the Kerr spacetime, is a
decreasing function of $a/M$, and the angular frequency for the
innermost orbit is an increasing function of
$a/M$\cite{SundararajanEtAlI}. The period of the oscillations as the
plunge is approached are in agreement with the analytically known frequencies at
the ISCO. 
It is tempting to infer that this higher pre-plunge frequency 
for $a/M=0.9$ as compared with $a/M=0.6$  accounts for the  greater antikick cancellation 
of the kick, just as the toy model with $\Omega=3$ exhibited greater antikick 
cancellation than did the smaller values of $\Omega$. We will argue that this temptation 
should be resisted.

\begin{figure}[h]
  \begin{center}
  \includegraphics[width=.7\textwidth ]{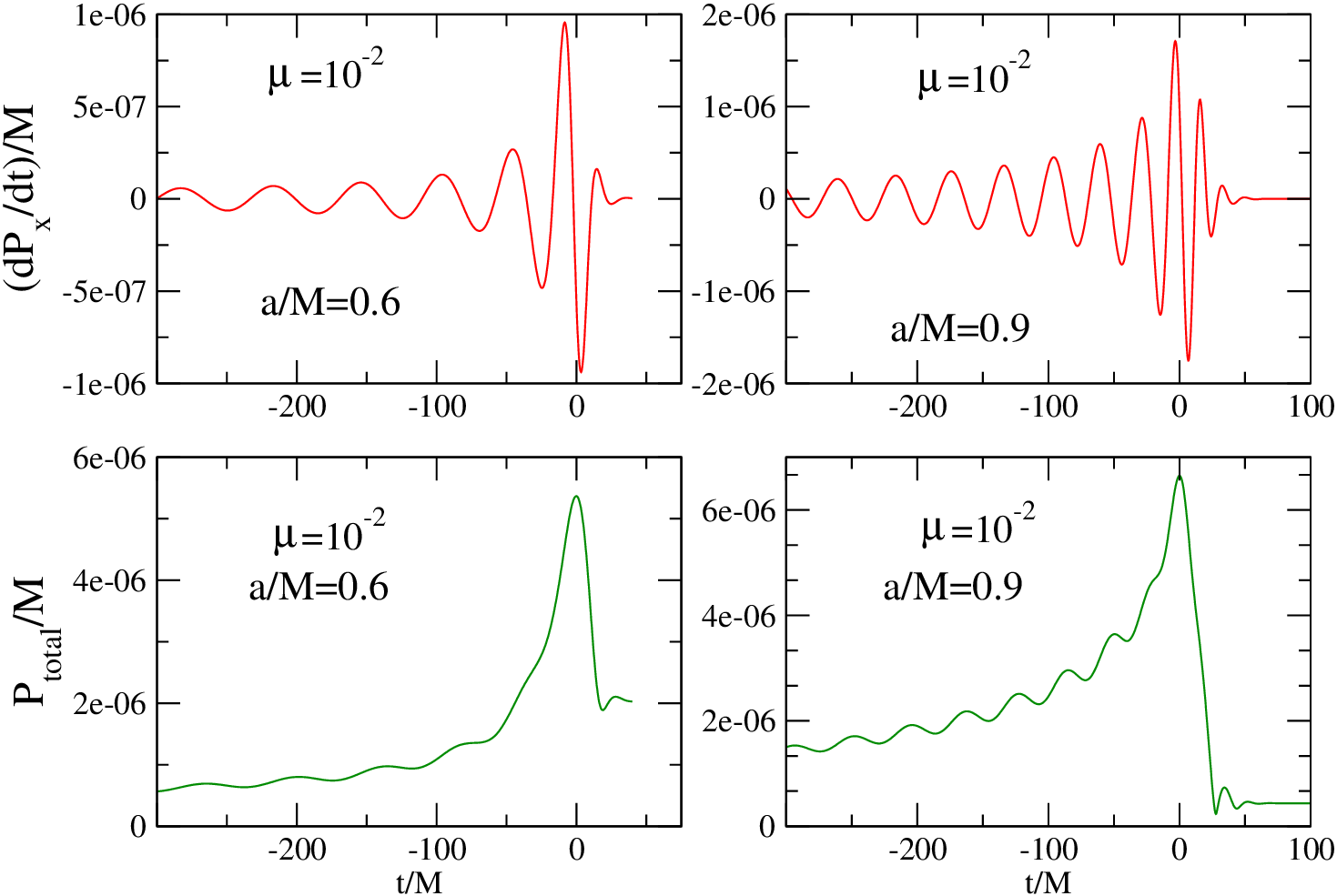}
  \caption{Computational results for mass ratio $\mu=10^{-2}$ in the case of 
spin parameters $a/M=0.6$ and $a/M=0.9$.}
  \label{fig:mu-2}
  \end{center}
  \end{figure}

Figure~\ref{fig:mu-2} shows the results for mass ratio $\mu=10^{-2}$
that correspond to the results 
for $\mu=10^{-4}$
in Fig.~\ref{fig:mu-4}. The most
notable difference in the momentum generation in the two cases is that
the pre-plunge evolution of the envelope is significantly faster for
$\mu=10^{-2}$ than it is for $\mu=10^{-4}$. If the antikick cancellation 
depended on how gradually the envelope of oscillations changes prior to 
the plunge then we would expect that the $\mu=10^{-2}$ results would exhibit 
\smallskip
weaker antikick cancellation than the $\mu=10^{-4}$ results. 
This turns out 
not to be the case; the fraction of the kick cancelled is roughly the same
for the two values of $\mu$.
The computed cancellation for these models, and also for $a/M=0.7$ and $a/M=0.8$,
are shown in Table~\ref{tab:mu-2}.
\begin{table}[h]
\begin{tabular}{|c|c|c||c|c|c|c|}
\hline
&
\multicolumn{3}{|c||}{mass ratio $\mu=10^{-2}$}
&\multicolumn{3}{|c|}{mass ratio $\mu=10^{-4}$}
\\
\hline
$a/M$& $P_{\rm max}$&$P_{\rm final}$& cancelled
& $P_{\rm max}$&$P_{\rm final}$& cancelled\\
\hline
\hline
0.6&$5.37\times10^{-6}$&$2.03\times10^{-6}$    &62\%&$7.4\times10^{-10}$&$2.7
\times10^{-10}$&64\%\\
0.7&$5.57\times10^{-6}$&$1.61\times10^{-6}$&71\%&$8.3\times10^{-10}$&$2.0\times10^{-10}$&76\%\\
0.8&    $6.46\times10^{-6}$& $5.92\times10^{-7}$ &91\%&$9.5\times10^{-10}$&$1.1\times10^{-10}$&88\%\\
0.9& $6.48\times10^{-6}$  &   $4.36\times10^{-7}$& 93\%&$1.2\times10^{-9}$&$3.2\times10^{-11}$&97\%\\
\hline
\end{tabular}
\caption{Momentum results for models with mass ratios $\mu=10^{-2}$ and
  $\mu=10^{-4}$. The values of $P_{\rm max}$ and $P_{\rm final}$ are the maximum and
  final value of the momentum radiated. The cancelled column,
  $1-(P_{\rm final}/P_{\rm max}) $, is the fraction of the maximum
  kick that is cancelled by the later antikick.\label{tab:mu-2}}
\end{table}

Particle perturbation theory tells us that the amount of energy
radiated during the preplunge inspiral, divided by the background
mass, scales as $\mu$. This is approximately confirmed by the results
in Table~\ref{tab:mu-2}.  The confirmation is only approximate because
the nature of the onset of plunge is determined by the join of the
quasicircular inspiral and the plunge, which represents dynamics that
does not scale with the same $\mu$ factor as the early inspiral. What
is particularly noteworthy is that the fraction of kick canceled by
the antikick is rather insensitive to $\mu$, and hence insensitive to
how slow the early inspiral is.

\subsection{Implications of the results}

If it is not the rate of change in the early slow inspiral that
governs the size of the antickick, then what {\em does} govern it?
The alternative explanation  is the steepness of the post-plunge envelope. That
is, the antikick cancellation of a large fraction of the kick depends
on the extent to which the post-plunge envelope obeys the slowness
condition.  
To test this explanation we return to the toy model of Eq.~(\ref{envelope}), 
but in the form
\begin{equation}\label{adjtoy}
  f(t)=\frac{e^{t/\tau_1}}
{1+e^{(t-1)/\tau_2}}\,,
\end{equation}
with an adjustable early timescale $\tau_1$ and late timescale
$\tau_2/[1-(\tau_2/\tau_1)]$. For this envelope function, we compute the
``momentum'' using the same integrals as in Eq.~(\ref{toyPs}). Table
\ref{tab:toy} gives the results of the peak and final ``momentum'' for
$\Omega=1.$ What is immediately striking about the results is that changes in the
early slow timescale do not have a large effect on either the peak
momentum or the final momentum, and that this is true for both values
of the late fast timescale. A change in the late fast timescale, however, has a
strong effect on the cancellation.

\begin{table}[h]
\begin{tabular}{|c|c|c||c|c|c|c|}
\hline
&
\multicolumn{3}{|c||}{$\tau_2 =1$}
&\multicolumn{3}{|c|}{$\tau_2=2$}
\\
\hline
$\tau_1$& $P_{\rm max}$&$P_{\rm final}$&cancelled
& $P_{\rm max}$&$P_{\rm final}$& cancelled\\
\hline
\hline
5&0.842&0.332    &61\%&0.671&0.0287&95.7\%\\
10&0.854&0.301&65\%&0.698&0.0259&96.3\%\\
15& 0.875& 0.291 &67\%&0.743&0.0251&96.6\%\\
30& 0.914  & 0.281& 69\%&0.821&0.0243&97.0\%\\
\hline
\end{tabular}
\caption{Momentum results for the toy model with the envelope function
defined in Eq.~(\ref{adjtoy}), and with $\Omega=1$.
\label{tab:toy}}
\end{table}

It is not ``early'' and ``late'' that are relevant here.  (This is
obvious in the fact that the momentum integrations can be run
backwards in time\cite{backwards}.)  What is relevant is that making the slow process
slower has little effect; making the fast process slower has a strong
effect. In the context of binary inspiral/merger this means that it is 
the  plunge and quasinormal ringing that are crucial to determining
the impact of the  antikick cancellation. This insight is potentially useful
in numerical relativity modeling. It suggests that lessons learned from 
the very late epoch of modeling are what is important to cancellation, and 
that models can be started very late. 

This insight about the impact of the speed of the late time process
has implications also for understanding the qualitative roots of
kick-antikick cancellation.  For particle perturbation computations of
radiated linear momentum reported in Ref.~\cite{SundararajanEtAlI},
the strength of the antikick is a rapidly increasing function of
$a/M$.  As exhibited in Fig.~5 of Ref.~\cite{SundararajanEtAlI}, the
antikick is very small for $a/M=-0.3$, and negligible for $a/M=-0.6$
and $-0.9$. (Negative values here indicate that the particle orbit is
retrograde.) The patterns of momentum generation indicate that with a 
decrease of $a/M$, the momentum generation cuts off more quickly at the start 
of plunge. A plausible speculation is that this more rapid cut off is related to 
the larger ISCO for smaller $a/M$. 

In the case of the inspiral/merger of comparable mass holes there is
evidence in Fig.~15 of Schnittman et al.\cite{SchnittmanEtAl} that the 
correlations noted in particle perturbation models also apply. In particular,
the antikick is most pronounced in the model in which the more massive hole
has spin aligned with the orbital angular momentum. Though there is no strict
meaning to an ISCO for binaries of comparable mass holes, this alignment of 
spins should play a role similar to that in EMRIs in governing the onset 
of the plunge-like epoch.

\begin{figure}[h]
  \begin{center}
  \includegraphics[width=.6\textwidth ]{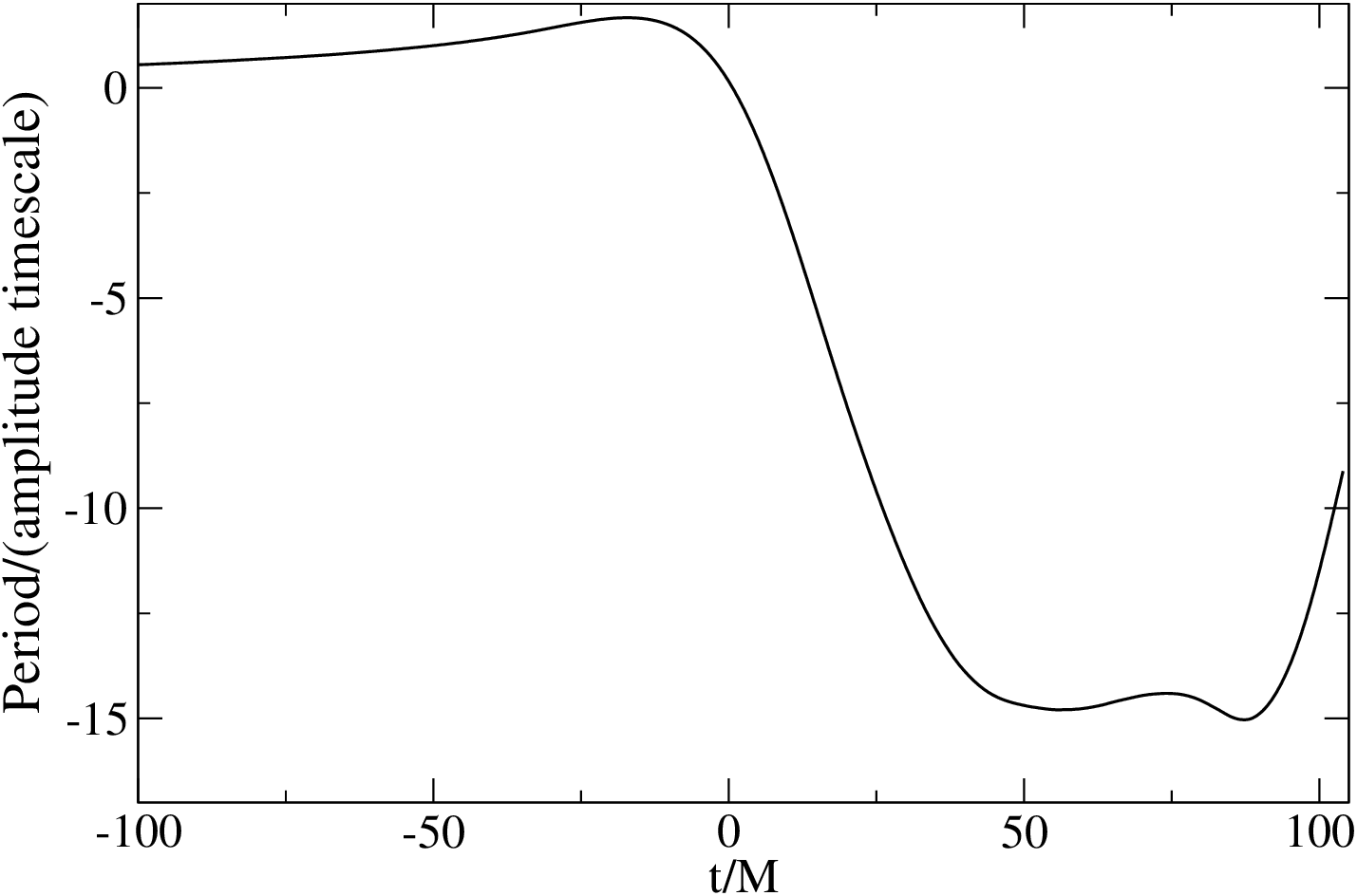}
  \caption{An indicator of ``slowness,'' the ratio of the oscillation
period to the timescale for change in the strength of the momentum emission,
for the particle-perturbation model with $\mu=10^{-4},$ and $a/M=0.9$.}
  \label{fig:periodtoscale}
  \end{center}
  \end{figure}

It is clear that the antikick cancellation is closely connected with
how gradually the intensity and period of momentum change. It is
natural, therefore, to seek a quantification of the ``slowness'' of
the momentum generation. The slowness is a function of time due to the
intensity of the momentum generation ($f(t)$ of Eq.~(\ref{LxLy})) and
the period ($T=2\pi/\Omega(t)$).  These are simple functions of time
both in the early gradual inspiral and in the late quasinormal
ringing, but cannot be simply characterized during the plunge.  The
period is particularly difficult to pin down during the plunge, since
the plunge -- the transition from quasicircular orbits to quasinormal
ringing -- typically takes only one oscillation, at least in the most interesting
cases, the particle parturbation models with large positive $a/M$.  As an indication of
the difficulty of quantifying slowness we have therefore used the
timescale for change in the envelope of the momentum oscillations
$\tau=\left(d\ln{f}/dt\right)^{-1}$. To get the period as a function
of time, a simple (and ultimately subjective) analytic fit was made to
the period for a few cycles before and after the plunge (i.e., the
peak of emission). Figure~\ref{fig:periodtoscale} shows the results
for the slowest case, that of $\mu=10^{-4}$ and $a/M=0.9$, the case for which
slowness should be most easily quantified. The figure
shows that the early inspiral is ``slow'' by this criterion. (The
period is much shorter than the timescale for change.) As the process
approaches the plunge, it becomes faster, but still satisfies the
criterion of being reasonably slow. However, at the plunge
(interpreted here as the steep descent and low portion of the curve in
Fig.~\ref{fig:periodtoscale}) the process seems badly to violate the
slowness criterion.  Though the slowness approximation explanation for
the systematics of the cancellation appears to give an excellent
accounting of all results, we see that it is not strictly valid. This
can be understood, at least partially, from the fact that when the
slowness approximation fails, the rate of generation of momentum has
already greatly decreased. 

It would of course be useful, or at least satisfying, to have a
quantification of slowness that is more meaningful than that exhibited
in Fig.~\ref{fig:periodtoscale}, but this is probably not possible, as
had already been suggested by Damour and
Gopakumar\cite{DamourGopakumar06}.  The greatest challenge is that a
quantification requires that we characterize the frequency through the
plunge. For Fig.~\ref{fig:periodtoscale} we have used the crude method
of noting zero crossings and fitting a simple function for
$\Omega(t)$. Much more sophisticated methods exist for separating
changes into those of amplitude and frequency, such as the normalized
Hilbert transform\cite{HHT}, but no
method can effect such a separation in the case that the period is
comparable to the timescale for change. A more promising approach may
be a semianalytic joining of the quasicircular inspiral to the
quasinormal ringing rooted in modeling of the EMRI or comparable mass
processes.

\section{The slowness approximation as an aid to model computations}
\label{sec:slowapp}

The analysis of the previous section can be used to reduce
the computational burden in running models, both with the particle
perturbation approximation, and with fully nonlinear gravity. The
underlying principle is that the slowness approximation is highly
justified in the very early stages of inspiral. By exploiting the
slowness approximation we can eliminate the need to carry out model
computations from very early times; models can be run starting at
relatively late times. Here we focus on using the approximation for
momentum computations.

\subsection{The $1/\Omega$ approximation}
The fundamental idea in using the slow approximation to replace early
orbits is to use the approximations in Eq.~(\ref{LofT}), ignoring the corrections
of order $T/\tau$. With corrections ignored, Eqs.~(\ref{LxLy}) and \ref{LofT}
give us
\begin{equation}\label{eq:slowapp}
  P_x(t)=\dot{P}_y(t)/\Omega(t)\quad\quad\quad  {P}_y(t)=-\dot{P}_x(t)/\Omega(t)\,.
\end{equation}
To illustrate the validity of this approximation we 
choose the example $\mu=10^{-2}$ and $a/M=0.6$. For the slow approximation 
this should be the most severe test, since the low mass ratio and the 
moderate value of $a/M$ both favor fast changes. 

  \begin{figure}[h]
  \begin{center}
  \includegraphics[width=.8\textwidth ]{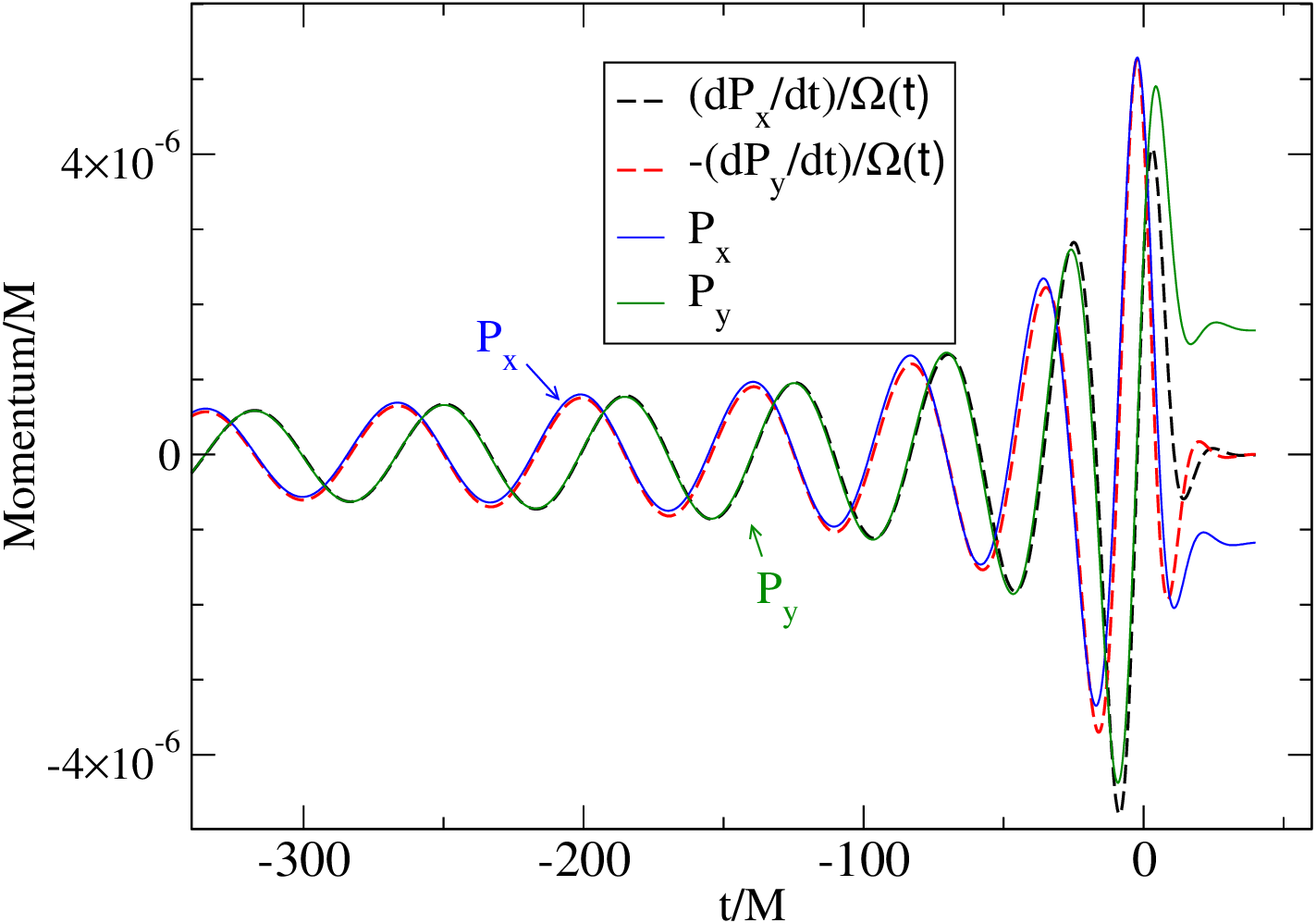}
  \caption{Momentum radiated during inspiral for a model with $\mu=10^{-2}$, $a/M=0.6$.
The solid curves are the computed momenta as a function of time. The dashed curves
are the approximation according to Eq.~(\ref{eq:slowapp}).
}
  \label{fig:slowapp}
  \end{center}
  \end{figure}

Figure~\ref{fig:slowapp} shows that even for this case the success of
the slow approximation is remarkable. The figure shows the comparison
of the two sides of each of the equalities in
Eq.~(\ref{eq:slowapp}). We see that before the plunge $ P_x(t)$ is
almost indistinguishable from $\dot{P}_y(t)/\Omega(t)$ and $ P_y(t)$
is almost indistinguishable from $-\dot{P}_x(t)/\Omega(t)$. The
momentum values, and their slow approximation values, begin to deviate
noticeably only around $t/M=20$, only slightly earlier than the
nominal start of plunge at $t/M=0$.

The usefulness of this approximation is clear. If we want to compute
the momentum radiated during an inspiral, we need not start at a time
long before the plunge. In fact, we need only start early enough so
that the inevitable numerical noise from the starting process has
subsided by the time the model is within a single oscillation of the
plunge!

\subsection{Integration from the peak}

We now describe a technique for finding the value of
$P_x$ that requires only the computation of $\dot{P}_x$
(and similarly for $P_y$). In this
approach we note that $P_x=0$ at a value of $t/M$ at which
$\dot{P}_y=0$. To the same order in the approximation, it is a time
at which $|\dot{P}_x|$ is a maximum. This means that we will get
results that are accurate to the order of the approximation if we
start integrating for $P_x$, with $P_x$ set to zero, at a time when
$\dot{P}_x$ is a maximum. An appropriately modified version of this
statement applies for the computation of $P_y$.

\begin{figure}[h]
  \begin{center}
  \includegraphics[width=.6\textwidth ]{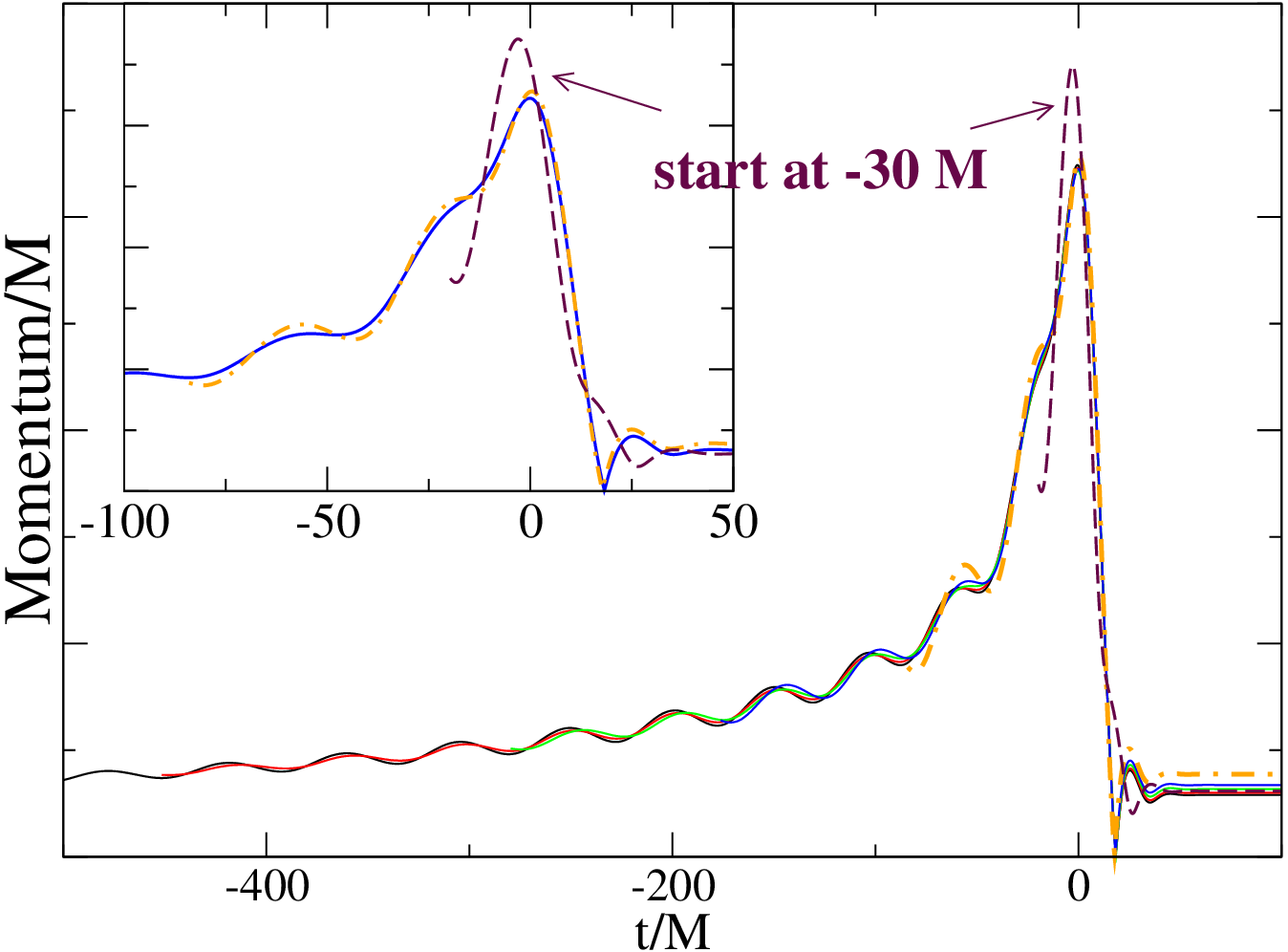}
  \caption{Momentum computed, as a function of time, using integration from 
the peak. For models with $\mu=10^{-2}$, $a/M=0.8$, results are shown 
for the total momentum computed using
integration of $P_x$ starting from a maximu of $\dot{P}_x$, 
and similarly for $P_y$. Plots are included starting approximately at times -700M, -500M,
-300M, -200M, -100M and -30M. The curve for integration started at -100M has
dots and dashes; the curve for integration starting at -30M
is shown with long dashes. Remarkably even the -100M curve, starting very near the plunge.
gives a good approximation for the momentum.
}
  \label{fig:IFP}
  \end{center}
  \end{figure}

In Fig.~\ref{fig:IFP} we show 
the results of this method used to compute the radiated momentum for
the model with $\mu=10^{-2}$ and $a/M=0.8$.  The curves give results for
integration starting at different epochs. Since the peak of the rate
of emission occurs around $t/M=0$, it is rather remarkable that
integration starting as late as $t/M=-80$, gives results that are the
same, within the accuracy of the modeling, with integration started around
$t/M=-700$. 

\section{Conclusions}\label{sec:conclusion}
The explanations and approximations in the previous section can be
considered to be based on understanding the inspiral radiation for
circular, equatorial binary inspiral with a model that is
rooted in the idea of a rotating beam of radiated momentum, whether 
in a particle perturbation computation or in numerical relativity 
computations with the fully nonlinear theory. For radiated momentum
confined to the orbital plane we have shown that an explanation of 
the kick/antikick cancellation follows from treating the change in 
the amplitude and frequency of the momentum components as slow compared
to the period of the oscillations.

In the early quasicircular inspiral, this slowness approximation is
rigorously true and leads to efficient methods for starting a
computation at late times, just before the plunge, while getting an
accurate value of the momentum radiated since the start of inspiral.

We have argued that no simple prescription exists for characterizing
the momentum during the plunge. 
Damour and Gopakumar\cite{DamourGopakumar06}, arguing similarly, pointed
out that this prevented an accurate estimate of radiated momentum based 
on the effective one body approach.

Though a simple prescription is ruled out, it is plausible that a
study of the plunge dynamics will provide guidelines that allow
explanations and approximations for antikick cancellations. This is
very likely in the particle perturbation approach, in which the
plunge dynamics is governed by orbits in the Kerr background.
The lessons learned from particle perturbation models may lead to 
a better understanding, or categorization of the antikick for the
inspiral/merger of comparable mass holes.

Most of the insights and methods above are specific to the case of
inspiral/merger in which the momentum is confined to the orbital
plane.  The general inspiral/merger, however, will not have this
equatorial symmetry. In the case of a particle perturbation model, the
particle orbit will be inclined to the equatorial plane and there will
be a component of radiation in the $z$ direction (parallel to the hole
angular momentum) as well as in the orbital $xy$ plane. The simple fit
to an oscillation (with a time changing frequency) will not suffice,
since at any epoch there will be a motion in the $z$ direction as well
as in the $x$ and $y$ directions. There will then be two time varying 
frequencies, each with its time varying amplitude. It is interesting to 
ask whether there are any insights or results in such cases analogous to 
those for the simple equatorial case. 

Initial numerical experiments suggest that it is possible to fit the
late pre-plunge inspiral to a two frequency model, at least in the
case that of particle perturbation results with only a small tilt out
of the equatorial plane. We intend to pursue this idea for
nonequatorial orbits both with particle perturbation results and with
results from numerical relativity. Of particular interest is an
objective procedure in which the two nonconstant frequencies are
extracted from the late pre-plunge data using the Hilbert-Huang
transformation \cite{HHT} a technique that is ideally suited to
extract sets of modes that have different, nonconstant frequencies.

\section*{Acknowledgments}

RHP gratefully acknowledges support of this work by NSF Grant 0554367,
and from the UTB Center for Gravitational Wave Astronomy.  This work was
supported at MIT by NASA Grant No.\ NNG05G105G and NSF Grant
PHY-0449884.  GK acknowledges research support from NSF Grants
PHY-0902026, CNS-0959382 and PHY-1016906, and hardware donations from
Sony and IBM.


\begin{thebibliography}{11}
\bibitem[{typ()}]{typo}
\bibinfo{note}{In Eq.~(5.2) of Ref.~\cite{SundararajanEtAlI} the reported
  maximum recoil for this case contains a misprint. The actual value of $v^{\rm
  peak}/c$ is $0.12(\mu/M)^2$ rather than $0.17(\mu/M)^2$. The value of $v^{\rm
  peak}/c$ shown in Fig.~5 of that paper is correct.}
\bibitem[{bac()}]{backwards}
\bibinfo{note}{The time reversal also requires, in principle, addititions to
  the momentum so that the $t=\infty$ value of $P_x,P_y$ are zero.}
\end{thebibliography}
\end{document}